\documentclass[prb,amsmath,amssymb,aps,twocolumn,superscriptaddress,longbibliography]{revtex4-2}
\usepackage{appendix}
\usepackage{bm}
\usepackage{graphicx}
\usepackage{epsfig}
\usepackage{epstopdf}
\usepackage{balance}
\usepackage[dvipsnames]{xcolor}
\usepackage{calc}
\usepackage{natbib}
\usepackage[colorlinks,
            linkcolor=blue,
            anchorcolor=blue,
            citecolor=blue,
            urlcolor=blue]{hyperref}
\usepackage{lipsum}
\usepackage[version=3]{mhchem} 

\usepackage{color}
\usepackage{soul}
\usepackage[etex=true,export]{adjustbox}
\usepackage{makecell}
\usepackage{multirow}
\usepackage{tabularx,multirow,array,diagbox}
\usepackage{adjustbox}
\usepackage{booktabs}
\makeatletter
\renewcommand{\maketag@@@}[1]{\hbox{\m@th\normalsize\normalfont#1}}%
\makeatother
\begin{document}
\title{Electrically tuned topology and magnetism in twisted bilayer MoTe$_2$ at $\nu_h=1$}

\author{Bohao Li}
\affiliation{School of Physics and Technology, Wuhan University, Wuhan 430072, China}
\author{Wen-Xuan Qiu}
\affiliation{School of Physics and Technology, Wuhan University, Wuhan 430072, China}
\author{Fengcheng Wu}
\email{wufcheng@whu.edu.cn}
\affiliation{School of Physics and Technology, Wuhan University, Wuhan 430072, China}

\begin{abstract}
We present a theoretical study of an interaction-driven quantum phase diagram of twisted bilayer MoTe$_2$ at hole filling factor $\nu_h=1$ as a function of twist angle $\theta$ and layer potential difference $V_z$, where $V_z$ is generated by an applied out-of-plane electric field.  At $V_z=0$, the phase diagram includes quantum anomalous Hall insulators in the intermediate $\theta$ regime and topologically trivial multiferroic states with coexisting ferroelectricity and magnetism in both small and large $\theta$ regimes. There can be two transitions from the quantum anomalous Hall insulator phase to topologically trivial out-of-plane ferromagnetic phase, and finally to in-plane 120$^\circ$ antiferromagnetic phase as $|V_z|$ increases, or a single transition without the intervening ferromagnetic phase. We show explicitly that the spin vector chirality of various 120$^\circ$ antiferromagnetic states can be electrically switched. We discuss the connection between the experimentally measured Curie-Weiss temperature and the low-temperature magnetic order based on an effective Heisenberg model with magnetic anisotropy.
\end{abstract}
\maketitle

\textit{Introduction.---} 
Coulomb interactions between electrons in topological flat bands can drive exotic quantum states of matter, as exemplified by fractional quantum Hall insulators in Landau levels.  Moir\'e systems with different combinations of layered van der Waals materials become a versatile platform to study nearly flat moir\'e bands with nontrivial topology \cite{Bistritzer2011,Cao2018,Cao2018a,Wu2019,Pan2020,Yahui2019,Serlin2020,Sharpe2019, Polshyn2020,Po2018,Po2019,Song2019,PhysRevB.99.155415,
PhysRevX.9.031021,Stepanov2021,Ledwith2020,xie2021fractional,Devakul2021,Zhang2021,Xie2022a,Devakul2022,Pan2022,Xie2022,abouelkomsan2022multiferroicity}. It was theoretically predicted \cite{Wu2019} that twisted transition metal dichalcogenide (TMD) homobilayers can host topological moir\'e bands that effectively realize the Kane-Mele model \cite{Kane2005,Kane2005a} for quantum spin Hall insulators. A series of recent experiments reported convincing observations of not only integer but also fractional quantum anomalous Hall insulators in twisted bilayer MoTe$_2$ ($t$MoTe$_2$) \cite{Anderson2023, Cai2023,Zeng2023,Park2023b,Xu2023}. Spectroscopic evidence of the integer quantum anomalous Hall insulators was also found in twisted bilayer WSe$_2$ \cite{Foutty2023}. These exciting discoveries open up many opportunities in condensed matter physics \cite{jain2023twist}, and attract active theoretical study on the nature of interaction-driven states in twisted TMD homobilayers at integer and fractional filling factors \cite{Wang2023,reddy2023,Qiu2023,dong2023a,wang2023topological,luo2023majorana,goldman2023,NM2023,song2023phase,liu2023gate,xu2023maximally,abouelkomsan2023band,yu2023fractional}.

\begin{figure}[t]    
\includegraphics[width=1\columnwidth]{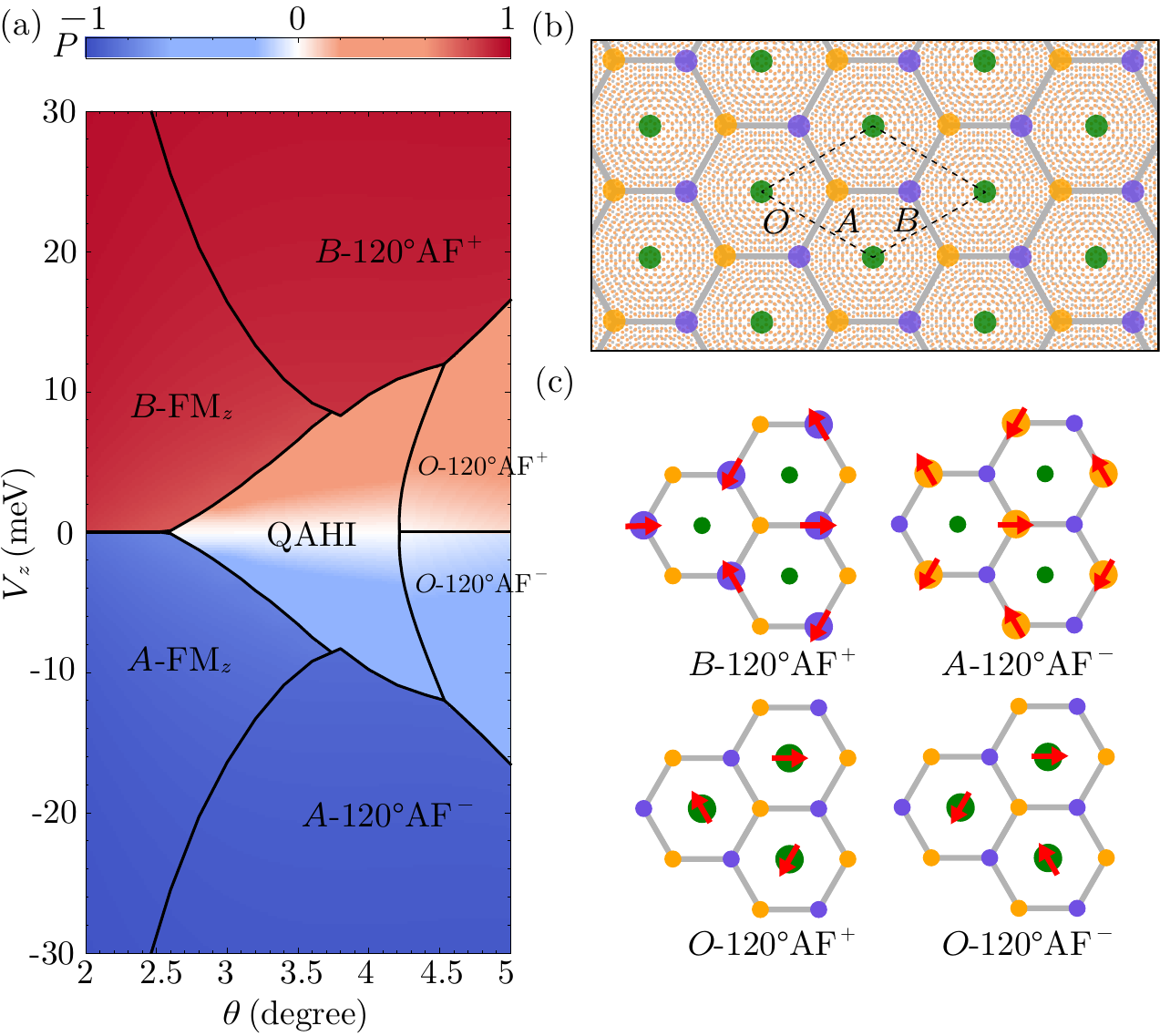}
    \caption{(a) Quantum phase diagram at $\nu_h =1$ as a function of $V_z$ and $\theta$. The color map represents the layer polarization $P$. (b) Moiré superlattices of $t$MoTe$_2$, where $O$, $A$, and $B$ are three high-symmetry sites. (c) Schematic illustration of the four 120$^\circ$ antiferromagnetic states with different spatial occupations and spin vector chiralities.} 
    \label{fig:1}
\end{figure}

In this Letter, we report the quantum phase diagram of $t$MoTe$_2$ at $\nu_h=1$ (i.e., one hole per moir\'e unit cell) tuned by twist angle $\theta$ and layer potential difference $V_z$, which is obtained by a mean-field study of an interacting continuum model in the plane-wave basis. The phase diagram shown in Fig.~\ref{fig:1}(a) hosts the quantum anomalous Hall insulator (QAHI) in the intermediate $\theta$ and small $|V_z|$ regime. The QAHI has spontaneous valley (out-of-plane spin) polarization, and the topology of the state arises from the winding of layer pseudospin in the moir\'e band. A finite $V_z$ potential induces layer polarization and therefore, drives phase transition from the QAHI state to topologically trivial magnetic states with holes primarily localized within one layer.  Our phase diagram is consistent with available experimental findings \cite{Anderson2023, Cai2023,Zeng2023,Park2023b,Xu2023} and related theoretical studies of the $\nu_h=1$ phase diagram based on different approaches \cite{Qiu2023, dong2023a, wang2023topological}. \textcolor{black}{Our main results} are summarized as follows. 
First, we present a global phase diagram using a self-consistent Hartree-Fock approximation without projecting to a few selected moir\'e bands. Details of the phase transitions from the QAHI to topological trivial magnetic insulators are revealed. There can be two transitions from the QAHI phase to the topologically trivial ferromagnetic phase with out-of-plane spin polarization, and finally to the in-plane $120^{\circ}$ antiferromagnetic ($120^{\circ}$AF) phase, or a single transition from the QAHI phase to the $120^{\circ}$AF phase, as $|V_z|$ increases. 
Second, the phase diagram clearly shows that the spin vector chirality of various $120^{\circ}$AF phases is controlled by the sign of $V_z$.
Third, the topologically trivial magnetic phases at large $|V_z|$ can be effectively described by a Heisenberg model with magnetic anisotropy on a triangular lattice. The Curie-Weiss temperature $T_{\text{cw}}$ relevant to the out-of-plane magnetic susceptibility $\chi_{zz}$ is given by $-3J_z/(2k_B)$, where $J_z$ is the out-of-plane spin coupling constant [see Eq.~\eqref{HS}]. The value of $J_z$ extracted from numerical results changes sign near the phase boundary between the out-of-plane ferromagnetic phase and in-plane antiferromagnetic phase.  Therefore, $T_{\text{cw}}$ as measured above the magnetic ordering temperature provides a strong indication of the magnetic order at low temperature.

\textit{Moiré Hamiltonian.---} 
The moir\'e Hamiltonian for valence band states in $t$MoTe$_2$ is given by \cite{Wu2019},
\begin{equation}
\begin{aligned}
\mathcal{H}_{\tau}=&
\begin{pmatrix}
\!-\frac{\hbar^2\left(\boldsymbol{k}-\tau\boldsymbol{\kappa}_{+}\right)^2}{2 m^*}+\Delta_{b}(\boldsymbol{r}) & \Delta_{T,\tau}(\boldsymbol{r}) \\
\Delta_{T,\tau}^{\dagger}(\boldsymbol{r}) & 
\!-\frac{\hbar^2\left(\boldsymbol{k}-\tau\boldsymbol{\kappa}_{-}\right)^2}{2 m^*}+\Delta_{t}(\boldsymbol{r})
\end{pmatrix}\\
&+\frac{1}{2}\begin{pmatrix}
    V_z & 0\\
    0 & -V_z
\end{pmatrix}\\
\end{aligned}
\label{Htau}
\end{equation}
where $\mathcal{H}_{\tau}$  is formulated in the $2\times2$ layer pseudospin space, $\tau=\pm$ is the spin (equivalent to valley) index, $\boldsymbol{k}=-i\partial_{\boldsymbol{r}}$ is the momentum operator, and $m^*= 0.62 m_e$ is the effective mass ($m_e$ is the electron rest mass). The vectors $\bm{\kappa}_{\pm}=\left[4\pi /(3 a_M)\right](-\sqrt{3}/2, \mp 1/2 )$ account for the rotation-induced momentum shift, where the moir\'e period $a_M$ is $a_0/\theta$ with $a_0= 3.472 $\AA ~being the monolayer lattice constant. The intralayer potentials $\Delta_{l}\left(\boldsymbol{r}\right)$ and the interlayer tunneling $\Delta_{T,\tau}\left(\boldsymbol{r}\right)$ are periodic functions of position $\bm{r}$,
\begin{equation}
\begin{aligned}
\Delta_{l}\left(\boldsymbol{r}\right) & =2 V \sum_{j=1,3,5} \cos \left(\boldsymbol{g}_j \cdot \boldsymbol{r} + l \psi\right), \\
\Delta_{T,\tau}\left(\boldsymbol{r}\right) & =w\left(1+e^{-i \tau \boldsymbol{g}_2 \cdot \boldsymbol{r}}+e^{-i\tau \boldsymbol{g}_3 \cdot \boldsymbol{r}}\right)
\end{aligned}
\end{equation}
where $l$ is $+1$ and $-1$, respectively, for the bottom ($b$) and top ($t$) layers,  $\boldsymbol{g}_j=\frac{4\pi}{\sqrt{3}a_M}(\cos\frac{\pi(j-1)}{3},\sin\frac{\pi(j-1)}{3})$ are the 
first-shell moiré reciprocal lattice vectors, and $(V,\psi, w)$ are model parameters.  

The diagonal terms $\pm V_z/2$ in $\mathcal{H}_{\tau}$ are generated by an out-of-plane electric displacement field. In the absence of the electric field ($V_z=0$), the moir\'e superlattices have $D_{3}$ point-group symmetry with $C_{3z}$ and $C_{2y}$ operations, where $C_{nj}$ is the $n$-fold rotaion about the $j$ axis. The $C_{2y}$ symmetry exchanges the two layers and is broken by a finite electric field ($V_z \neq 0$). 
A sufficiently large potential difference $V_z$ tends to polarize low-energy carriers into a single layer and effectively decouples the two layers.

The topology of moir\'e bands is characterized by the valley contrast Chern numbers $\mathcal{C}_{n,\tau}$, where $n$ is the band index. Due to time-reversal symmetry, $\mathcal{C}_{n,\tau}=-\mathcal{C}_{n,-\tau}$. The value of $\mathcal{C}_{n,\tau}$ depends on the model parameters \cite{Pan2020}. In this work, we focus on the parameter regime where $\mathcal{C}_{1,\tau}$ of the first moir\'e valence band is nontrivial at $V_z=0$. In this case, the first moir\'e band in each valley can be understood as a coherent superposition of two states that are, respectively, localized at $A$ and $B$ sites of the moir\'e superlattice [see Fig.~\ref{fig:1}(b)] and polarized to opposite layers \cite{Wu2019}. The $A$ and $B$ sites form a buckled honeycomb lattice since the two sites are also associated with different layers. The out-of-plane electric field produces a staggered potential on this honeycomb lattice. Therefore, the potential $V_z$ can drive the first moir\'e band from a layer-coherent state on a honeycomb lattice to a layer-polarized state on a triangular lattice, and further generate a  transition from a topological band ($\mathcal{C}_{1,\tau} \neq 0$) to a topologically trivial band ($\mathcal{C}_{1,\tau} = 0$).  

\begin{figure}[t]
    \includegraphics[width=1\columnwidth]{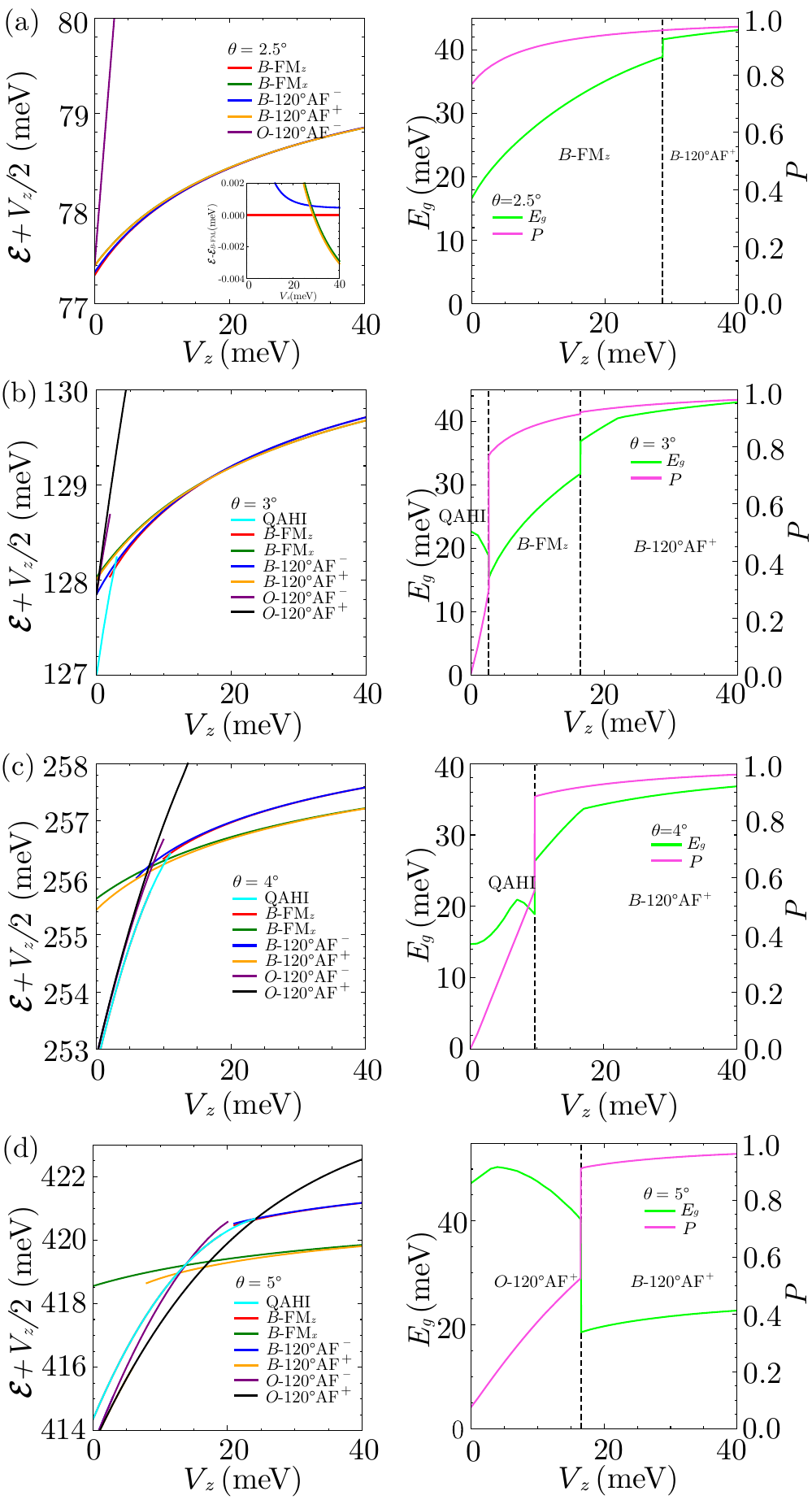}
    \caption{(a)-(d) Left panel:  $\mathcal{E}+V_z/2$ of various competing states at $\nu_h=1$ as a function of $V_z$, where $\mathcal{E}$ is energy per moir\'e unit cell. The inset in (a) plots the relative energy $\mathcal{E}-\mathcal{E}_{B-\text{FM}_z}$.   Right panel: charge gap $E_g$ and layer polarization $P$ in the $\nu_h=1$ ground state as a function of $V_z$.  The vertical black dashed lines mark phase transitions. Four representative values of $\theta$ are (a) $2.5^{\circ}$,(b) $3^{\circ}$,(c) $4^{\circ}$, and (d) $5^{\circ}$.}
    \label{fig:2}
\end{figure}

\textit{Coulomb interaction.---} 
To study many-body physics, we construct the full Hamiltonian $\hat{H}=\hat{H}_1+\hat{H}_{2}$, where $\hat{H}_1$ and $\hat{H}_2$ describe, respectively, the single-particle and interacting Hamiltonians. In the second quantization formalism, $\hat{H}_1$ is expressed as
\begin{equation}
\begin{aligned}
\hat{H}_1&=\sum_{\bm{k},\bm{k}'}\sum_{l,l'}\sum_{\tau} h_{\bm{k} l,\bm{k}'l'}^{(\tau)} c_{\bm{k},l,\tau}^\dagger c_{\bm{k}',l',\tau}\\
&=-\sum_{\bm{k},\bm{k}'}\sum_{l,l'}\sum_{\tau} [h^{(\tau)}]^{\intercal}_{\bm{k} l,\bm{k}'l'} b_{\bm{k},l,\tau}^\dagger b_{\bm{k},l',\tau},
\end{aligned}
\label{H1}
\end{equation}
where $c_{\bm{k},l,\tau}^\dagger$ ($c_{\bm{k}',l',\tau}$) is the electron creation (annihilation) operator of a plane-wave state with momentum $\bm{k}$ ($\bm{k}'$) in layer $l$ ($l'$) and valley $\tau$, and $h^{(\tau)}$ is the matrix representation of $\mathcal{H}_{\tau}$ in the plane wave basis. Since we study holes doped to the valence bands, it is more convenient to use the hole operator $b_{\bm{k},l,\tau}=c_{\bm{k},l,\tau}^\dagger$. In the second line of Eq.~\eqref{H1}, a constant term is dropped. 

The hole-hole Coulomb interaction is described by
 \begin{equation}
\hat{H}_{2}=\frac{1}{2 A} \sum_{\boldsymbol{k}, \boldsymbol{k}^{\prime}, \boldsymbol{q}} \sum_{l,l',\tau,\tau'} V(\boldsymbol{q}) b_{\boldsymbol{k+q},l,\tau }^{\dagger} b_{\boldsymbol{k^{\prime}-q},l^{\prime},\tau^{\prime}}^{\dagger} b_{\boldsymbol{k}^{\prime},l^{\prime},\tau^{\prime} } b_{\boldsymbol{k},l,\tau },
\end{equation}
where $A$ is the area of the system. Here we use the gate-screened Coulomb potential  $V(\bm{q})=2\pi e^2\tanh(|\bm{q}|d)/(\epsilon |\bm{q}|)$, where $\epsilon$ is the dielectric constant and $d$ is the gate-to-sample distance. The system has a weak dependence on $d$ for the typical case of $d \gg a_M$. We take $d$ to be 20 nm in our study. The full Hamiltonian $\hat{H}$ respects the point-group symmetry of the system, the time-reversal symmetry, and the valley (spin) U(1) symmetry.

\textit{Phase diagram.---} 
We study the interaction-driven quantum phase diagram as a function of $\theta$ and $V_z$ at the hole filling factor $\nu_h=1$. Here $\nu_h=n_b+n_t$, where $n_l=\frac{1}{N}\sum_{\boldsymbol{k},\tau} \left\langle b^{\dagger}_{\boldsymbol{k},l,\tau}b_{\boldsymbol{k},l,\tau}\right\rangle$ is the number of holes per moir\'e unit cell in layer $l$ and $N$ is the number of cells.
We take model parameters to be $V = 10$ meV, $\psi = -89.6^\circ$,  $w = -8.5$ meV, and $\epsilon=15$ based on our related study in Ref.~\onlinecite{Qiu2023}.
This previous work studied the $V_z=0$ phase diagram based on a three-orbital model with the Hamiltonian projected to the first three moir\'e bands \cite{Qiu2023}, where the above set of model parameters was found to correctly capture the $\nu_h=1$ QAHI experimentally observed at $\theta$ around $4^{\circ}$.
We note that low-energy holes are confined to the $B$ ($A$) site in the $b$ ($t$) layer for $\psi = -89.6^\circ$.

\begin{figure}[t]
    \includegraphics[width=0.9\columnwidth]{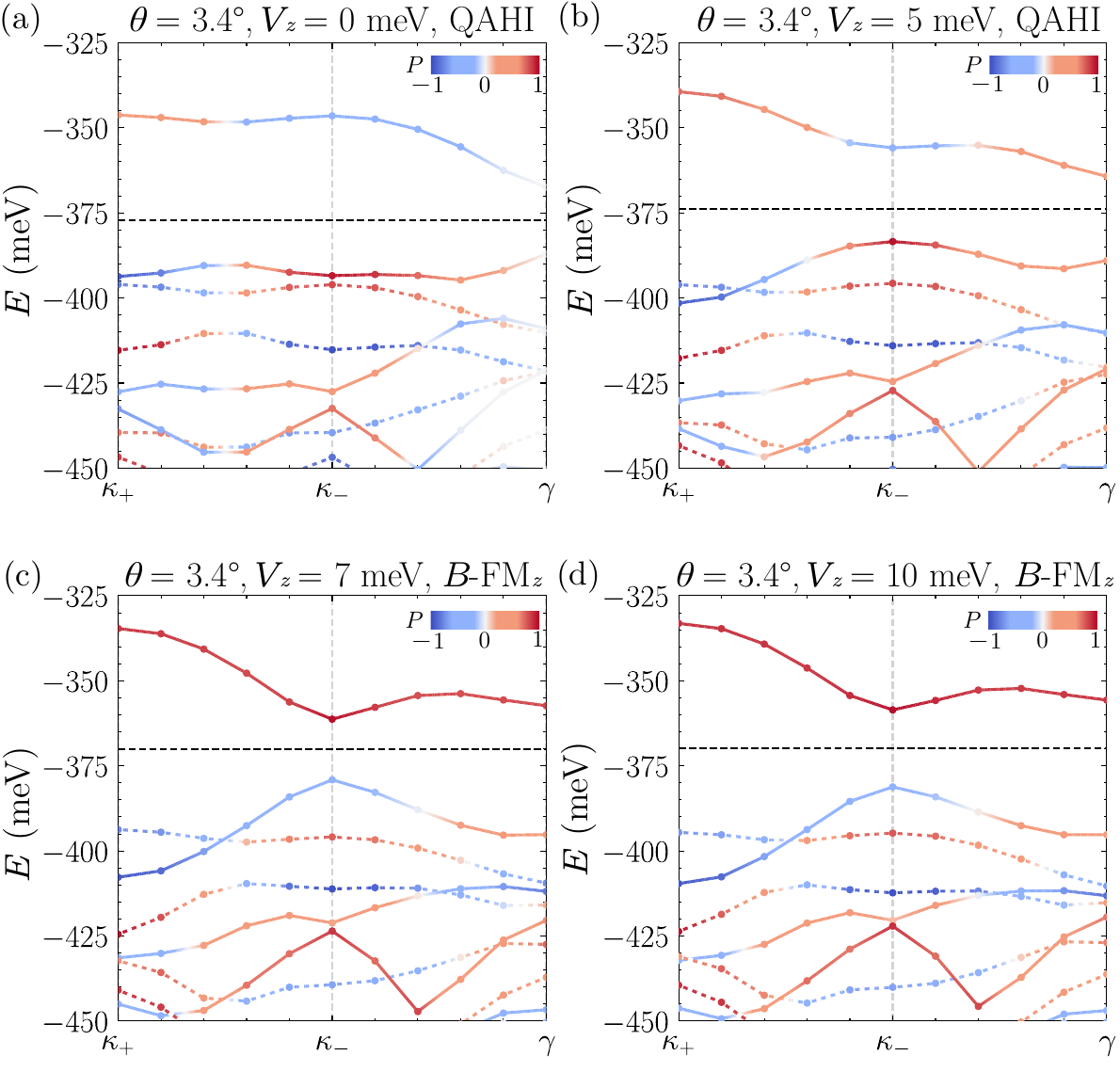}
    \caption{(a)-(d) The $\nu_h=1$ mean-field band structure at $\theta=3.4^{\circ}$ for different $V_z$. The band structure is presented in the basis defined by $c^{\dagger}_{\bm k, l, \tau}$ and $c_{\bm k, l, \tau}$ operators. The solid (dashed) lines plot bands with $\tau=+ (-)$. The color encodes the layer polarization of Bloch states. The middle of the interaction-induced gap is marked by the horizontal black dashed line. The band structure describes the QAHI state in (a) and (b), but the $B$-FM$_z$ state in (c) and  (d). The layer polarization of the unoccupied band above the chemical potential varies dramatically over the moir\'e Brillouin zone in the QAHI state and indicates the winding of layer pseudospin, but barely changes in the $B$-FM$_z$ state.}
    \label{fig:3}
\end{figure}

We perform a mean-field study of the Hamiltonian $\hat{H}$ using Hartree-Fock approximation in the plane-wave basis \textcolor{black}{following the procedure developed in Ref.~\onlinecite{Pan2022}} and present the phase diagram at $\nu_h=1$ in Fig.~\ref{fig:1}(a), which includes several magnetic phases as we will describe. The phase diagram is obtained by comparing the energies of multiple competing states, as illustrated in Fig.~\ref{fig:2}. The color map of Fig.~\ref{fig:1}(a) encodes the layer polarization $P=n_b-n_t$. Here $P$ of a given state is related to its energy per unit cell $\mathcal{E}$ by  $P=-2\partial \mathcal{E}/\partial V_z$ following the Hellmann–Feynman theorem. 

We start by describing the $\theta$-dependence of the phase diagram in Fig.~\ref{fig:1}(a) at $V_z=0$. (1) For $\theta<2.6^{\circ}$, there are two types of ground states labeled as $A$-FM$_z$ and $B$-FM$_z$, which have (spontaneous) opposite layer polarization and ferromagnetism along the out-of-plane $z$ direction (FM$_z$). Holes in $A$-FM$_z$ ($B$-FM$_z$) reside primarily in $A$ ($B$) sites and are spontaneously polarized to the $t$ ($b$) layer.  The FM$_z$ order results from spontaneous valley polarization.
\textcolor{black}{These states can be understood as charge density waves (generalized Wigner crystals) driven by the Coulomb repulsion between $A$ and $B$ sublattices on the honeycomb lattice \cite{abouelkomsan2022multiferroicity,Qiu2023} and belong to type-I multiferroics \cite{khomskii2009classifying}, where ferroelectricity and magnetism are nearly independent.} (2) For $\theta \in (2.6^{\circ}, 4.2^{\circ})$, the ground state realizes QAHI, which has spontaneous valley polarization but no layer polarization. The QAHI carries a quantized total Chern number $\mathcal{C}$ with $|\mathcal{C}|=1$, which is a result of the valley polarization and the band topology, as demonstrated by the mean-field band structure in Fig.~\ref{fig:3}(a). (3) For $\theta \in (4.2^{\circ},5.0^{\circ})$, there are also two types of ground states labeled as $O$-$120^{\circ}$AF$^{\pm}$, which are intervalley coherent states with opposite valley ordering wave vector \cite{Pan2020,Qiu2023}. In real space, the $O$-$120^{\circ}$AF$^{\pm}$ states have $120^{\circ}$ AF order developed mainly on $O$ sites [see Fig.~\ref{fig:1}(c)], which leads to an enlarged $\sqrt{3}\times\sqrt{3}$ magnetic unit cell.  
\textcolor{black}{The $O$-$120^{\circ}$AF$^{\pm}$ phases are distinguished by the spin vector chirality $\chi=(\bm{S}_{\bm{R}}\times\bm{S}_{\bm{R}+a_{M}\hat{y}})\cdot\hat{z}$, which is the cross product between spin vectors located at two neighboring $O$ sites. $\chi$ is $\pm 1$ for $O$-$120^{\circ}$AF$^{\pm}$, as illustrated in Fig.~\ref{fig:1}(c).}
\textcolor{black}{
Since $\chi$ changes sign under $C_{2y}$ rotation, the $120^{\circ}$ AF$^{\pm}$ order spontaneously breaks the $C_{2y}$ symmetry and results in spontaneous layer polarization $P$ that is locked to $\chi$, which is confirmed by numerical results in Fig.~\ref{fig:2}(d). Therefore, the $O$-$120^{\circ}$AF$^{\pm}$  states are type-II multiferroics \cite{Cheong2007,khomskii2009classifying,Song2022}, where ferroelectricity is generated by the magnetic order.}
Finally, we note that the phase diagram at $V_z=0$ in Fig.~\ref{fig:1}(a) is quantitatively consistent with that obtained from the three-orbital model \cite{Qiu2023}.

We now turn to the $V_z$-dependence of the phase diagram in Fig.~\ref{fig:1}(a), which is symmetric with respect to the $V_z=0$ line in the sense that states at $\pm V_z$  are related by the $C_{2y}$ rotation. (1) For $\theta<2.6^{\circ}$, an infinitesimal $V_z$ potential splits the degeneracy between $A$-FM$_z$ and $B$-FM$_z$ states due to their opposite layer polarizations. A positive (negative) $V_z$ potential favors the $B$-FM$_z$ ($A$-FM$_z$) state, and can further drive a transition into the $B$-$120^{\circ}$AF$^+$ ($A$-$120^{\circ}$AF$^-$) state at a critical $V_z$, where the magnetic order at $B$ ($A$) sites of the $b$ ($t$) layer changes from FM$_z$ to  $120^{\circ}$AF with positive (negative) spin vector chirality. (2) For $\theta \in (2.6^{\circ}, 4.2^{\circ})$, the QAHI state is robust up to some critical values of $V_z$. There are two ranges of $\theta$. 
In the first range of $\theta \in (2.6^{\circ}, 3.7^{\circ})$, there are transitions from the  QAHI state to the $B$-FM$_z$ ($A$-FM$_z$) state, and finally to the $B$-$120^{\circ}$AF$^+$ ($A$-$120^{\circ}$AF$^-$) state as $|V_z|$ increases. By contrast, there is a single transition from the  QAHI state to the $B$-$120^{\circ}$AF$^+$ ($A$-$120^{\circ}$AF$^-$) state at a critical $V_z$ for the second range $\theta \in (3.7^{\circ}, 4.2^{\circ})$.
(3) For $\theta \in (4.2^{\circ},5.0^{\circ})$, an infinitesimal $V_z$ potential again splits the degeneracy between $O$-$120^{\circ}$AF$^{\pm}$ states because of the ferroelectricity. A positive (negative) $V_z$ potential stabilizes the $O$-$120^{\circ}$AF$^{+}$ ($O$-$120^{\circ}$AF$^{-}$) state, and therefore, the spin vector chirality $\chi$ is controlled by the electric field. The $V_z$ potential drives the $O$-$120^{\circ}$AF$^{+}$ ($O$-$120^{\circ}$AF$^{-}$) state to the reentrant QAHI state, and finally to the $B$-$120^{\circ}$AF$^+$ ($A$-$120^{\circ}$AF$^-$) state for $\theta \in (4.2^{\circ},4.5^{\circ})$, but generates a single transition from the $O$-$120^{\circ}$AF$^{+}$ ($O$-$120^{\circ}$AF$^{-}$) state to the $B$-$120^{\circ}$AF$^+$ ($A$-$120^{\circ}$AF$^-$) state for $\theta \in (4.5^{\circ},5.0^{\circ})$.

The phase boundaries in Fig.~\ref{fig:1}(a), as determined by energy competition of various states (Fig.~\ref{fig:2}), mark first-order phase transitions. Physical quantities, such as charge gap $E_g$ and layer polarization $P$, generally have discontinuities across the phase boundaries, as shown in the right panel of Fig.~\ref{fig:2}. For example, $P$ jumps and changes sign across the $V_z=0$ phase boundary that separates the $O$-$120^{\circ}$AF$^{\pm}$ states. The QAHI and $B$-FM$_z$ ($A$-FM$_z$) states are distinguished by the Chern number $\mathcal{C}$, but they have the same symmetry breaking pattern at $V_z \neq 0$.  We numerically find that the transition between the QAHI and $B$-FM$_z$ ($A$-FM$_z$) states is first order; the charge gap has a dip at the transition, but does not need to fully close, as plotted in Fig.~\ref{fig:2}(b).

\textcolor{black}
{ 
We discuss our work in the context of literature. Previous studies of the $\nu_h=1$ phase diagram in $t$MoTe$_2$ projected interactions onto a few selected non-interacting moir\'e bands \cite{abouelkomsan2022multiferroicity, Qiu2023, dong2023a, wang2023topological}, while our calculation is performed in the plane-wave basis without the projection. Our results verify that the two approaches generate largely consistent phase diagrams, particularly on the field-tuned transition from the QAHI phase to topologically trivial magnetic phases \cite{dong2023a, wang2023topological}.  In addition, we reveal the \textit{real-space} pattern of different phases. For example, the $O$-$120^{\circ}$AF$^{+}$ and $B$-$120^{\circ}$AF$^{+}$ phases are both intervalley coherent states, but are distinguished by the real-space charge distribution.
}

\begin{figure}[t]
    \includegraphics[width=0.9\columnwidth]{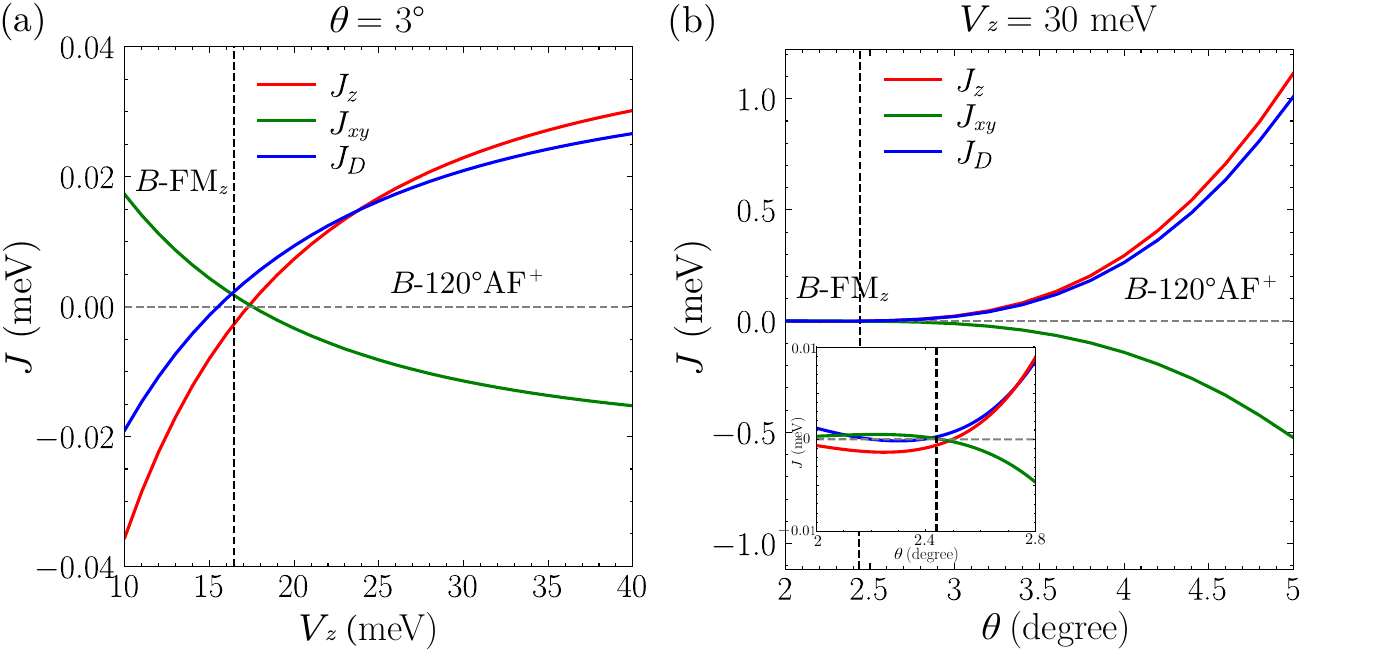}
    \caption{Numerical values of $J_z,J_{xy}$, and $J_D$ as functions of $V_z$ at $\theta=$3° in (a), and as  functions of $\theta$ at $V_z$=30 meV in (b). The vertical black dashed lines separate the $B$-FM$_z$ and $B$-$120^{\circ}$AF$^+$ phases. The inset in (b) is a zoom-in plot.}
    \label{fig:4}
\end{figure}

\textit{Effective Heisenberg model.---} 
In the phases with large layer polarization, the doped holes primarily reside at the $B$ ($A$) sites in the $b$ ($t$) layer, and the physics is then captured by a one-orbital Hubbard model on a triangular lattice \cite{Wu2018}. At $\nu_h=1$, the Hubbard model can be further mapped to an effective Heisenberg model. Without loss of generality, we focus on the $B$-FM$_z$ and $B$-120$^{\circ}$AF$^+$ phases at $V_z>0$, where the Heisenberg model can be parameterized as
\begin{equation}
\begin{aligned}
H_S = &J_z\sum_{\boldsymbol{R},i=1,3,5}S_{\boldsymbol{R}}^{z}S_{\boldsymbol{R}+\boldsymbol{\delta}_i}^{z}\\&
+J_{xy}\sum_{\boldsymbol{R},i=1,3,5}(S_{\boldsymbol{R}}^{x}S_{\boldsymbol{R}+\boldsymbol{\delta}_i}^{x}+S_{\boldsymbol{R}}^{y}S_{\boldsymbol{R}+\boldsymbol{\delta}_i}^{y})\\&
+J_D\sum_{\boldsymbol{R},i=1,3,5}(S_{\boldsymbol{R}}^{x}S_{\boldsymbol{R}+\boldsymbol{\delta}_i}^{y}-S_{\boldsymbol{R}}^{y}S_{\boldsymbol{R}+\boldsymbol{\delta}_i}^{x}),
\end{aligned}
\label{HS}
\end{equation}
where $\bm{S}_{\boldsymbol{R}}$ represents a spin-$1/2$ operator at a $B$ site, and $\boldsymbol{\delta}_i=a_M(\cos\frac{\pi (2i-1)}{6},\sin\frac{\pi (2i-1)}{6})$ connects nearest-neighbor $B$ sites. We include three types of nearest-neighbor spin interactions allowed by the spin (valley) U(1) symmetry, where the first two terms of $H_S$ are spin-exchange interactions in the XXZ Heisenberg model, while the last term is an effective Dzyaloshinskii-Moriya (DM) interaction. The coupling constants $J_z$, $J_{xy}$, and $J_D$ can be extracted from the mean-field energies of the following competing states of the continuum Hamiltonian $\hat{H}$: $B$-FM$_z$, $B$-FM$_x$, and $B$-120$^{\circ}$AF$^{\pm}$, which have, respectively, out-of-plane ferromagnetism, in-plane ferromagnetism, and 120$^{\circ}$ antiferromagnetism at $B$ sites. We note that the DM interaction favors noncollinear in-plane magnetism and generates the energy difference between $B$-120$^{\circ}$AF$^{\pm}$ states with opposite spin vector chiralities. The values of spin coupling constants obtained from the fitting are shown in Fig.~\ref{fig:4}. One important observation is that the sign of $J_z$, $J_{xy}$, and $J_D$ changes {\it near} (but not exactly at) the phase boundary between the $B$-FM$_z$ and $B$-$120^{\circ}$AF$^+$ phases. In the  $B$-FM$_z$ phase, $J_z<0$ is ferromagnetic, but $J_{xy}>0$ is antiferromagnetic. This is reversed in the $B$-$120^{\circ}$AF$^+$ state away from the phase boundary, where $J_z>0$ is antiferromagnetic and $J_{xy}<0$ is ferromagnetic. 
Another observation is that the values of the coupling constants grow by orders of magnitude as $\theta$ increases from $2^{\circ}$ to $5^{\circ}$ for a fixed large $|V_z|$, as shown in Fig.~\ref{fig:4}(b). This is expected since electron hoppings on the effective triangular lattice grow exponentially with decreasing $a_M$ (increasing $\theta$).

In the limit of $V_z \rightarrow +\infty$, the $t$ layer in the system can be neglected; the valley-dependent momentum shifts $\tau \bm{\kappa}_+$ of the $b$ layer in the moir\'e Hamiltonian $\mathcal{H}_\tau$ can be gauged away, which leads to an emergent valley (spin) SU(2) symmetry. A corresponding gauge transformation can be applied to the spin model $H_s$, which leaves the $J_z$ term invariant but results in an SU(2) symmetric Heisenberg model for  $V_z \rightarrow +\infty$.  The coupling constants in $H_S$ with this hidden SU(2) symmetry satisfy the constraint $J_{xy}/J_z=-1/2$ and $J_D/J_z=+\sqrt{3}/2$. Our numerical values of the coupling constants closely follow this constraint at large $V_z$.

In theory, there are competing magnetic interactions, such as antiferromagnetic superexchange and ferromagnetic intersite Hund's interaction \cite{Naichao_competing2021,Morales_nonlocal2022}. Therefore, both ferromagnetic and antiferromagnetic states are possible. In experiments, the magnetic interactions can be determined by measuring the magnetic susceptibility through the Curie-Weiss (CW) behavior \cite{Tang2020}. Due to the spin-valley locking in TMDs, spins can be coupled to an out-of-plane magnetic field $B_z$, but have negligible couplings to an in-plane magnetic field.  Therefore, available optical experiments based on magnetic circular dichroism \cite{Tang2020} can only measure the out-of-plane magnetic susceptibility $\chi_{zz} = \lim_{B_z \to 0} M_z/B_z$, where $M_z=\frac{1}{N}\sum_{\bm R} \langle S_{\bm R}^{z}\rangle$. Above the magnetic ordering temperature, the CW law indicates that $\chi_{zz} \propto 1/(T-T_{\text{cw}})$, where $T$ is the temperature and $T_{\text{cw}}$ is the CW temperature.  Using the effective spin model $H_S$ in Eq.~\eqref{HS}, we find that $k_B T_{\text{cw}}=-3J_z/2$. Therefore, the experimentally measured CW temperature $T_{\text{cw}}$ depends only on $J_z$, but not on the in-plane coupling constants $J_{xy}$ and $J_D$. Based on our numerical results of $J_z$ in Fig.~\ref{fig:4}, we find that the sign of $T_{\text{cw}}$ indeed provides a strong indication of the magnetic ordering at zero temperature.

\textit{Discussions.---} 
In summary, we present a rich phase diagram with a plethora of topological and topologically trivial magnetic phases that are tuned by $\theta$ and $V_z$.  
A recent experimental study with both transport and optical measurement revealed the field-induced two transitions from the QAHI phase to the  FM$_z$ phase, and finally to the AF phase at $\nu_h=1$ \cite{Park2023b}. Our phase diagram not only captures these two transitions driven by $V_z$, but also demonstrates the possibility of a single transition without the intervening FM$_z$ phase. Moreover, our analysis based on the effective spin model $H_S$ provides the theoretical mechanism for why the out-of-plane magnetic susceptibility measured optically by magnetic circular dichroism above the ordering temperature can indicate the low-temperature magnetic order. 
\textcolor{black}{
We view our results as qualitative instead of quantitative since complications such as lattice relaxation effects \cite{Wang2023,reddy2023} are not fully taken into account. Nevertheless, given the consistency with available experiments \cite{Anderson2023, Cai2023,Zeng2023,Park2023b,Xu2023}, we expect our phase diagram to be qualitatively correct.
We employ mean-field theory, while more exotic phases not captured by Hartree-Fock approximation, such as spin liquid phases\cite{PhysRevX.10.021042}, could also appear in the phase diagram.}

\textit{Acknowledgments.---} 
This work is supported by National Key Research and Development Program of China (Grants No. 2022YFA1402401 and No. 2021YFA1401300), National Natural Science Foundation of China (Grant No. 12274333), and start-up funding of Wuhan University. The numerical calculations in this paper have been performed on the supercomputing system in the Supercomputing Center of Wuhan University.

\bibliography{ref}
\end{document}